\documentclass[aip,jcp,reprint,amsmath]{revtex4-1}

\usepackage{bm}
\usepackage{natmove}
\usepackage{graphicx}
\usepackage{xcolor}
\usepackage[update,prepend]{epstopdf}
\begin{document}
\title{{Structure and phase behavior of polymer-linked colloidal gels}}

\author{Michael P. Howard}
\affiliation{McKetta Department of Chemical Engineering, University of Texas at Austin, Austin, Texas 78712, USA}

\author{Ryan B. Jadrich}
\affiliation{McKetta Department of Chemical Engineering, University of Texas at Austin, Austin, Texas 78712, USA}
\affiliation{Theoretical Division, Los Alamos National Laboratory, Los Alamos, New Mexico 87545, USA}
\affiliation{Center for Nonlinear Studies, Los Alamos National Laboratory, Los Alamos, New Mexico 87545, USA}

\author{Beth A. Lindquist}
\affiliation{McKetta Department of Chemical Engineering, University of Texas at Austin, Austin, Texas 78712, USA}
\affiliation{Theoretical Division, Los Alamos National Laboratory, Los Alamos, New Mexico 87545, USA}

\author{Fardin Khabaz}
\affiliation{McKetta Department of Chemical Engineering, University of Texas at Austin, Austin, Texas 78712, USA}

\author{Roger T. Bonnecaze}
\affiliation{McKetta Department of Chemical Engineering, University of Texas at Austin, Austin, Texas 78712, USA}

\author{Delia J. Milliron}
\affiliation{McKetta Department of Chemical Engineering, University of Texas at Austin, Austin, Texas 78712, USA}

\author{Thomas M. Truskett}
\email{truskett@che.utexas.edu}
\affiliation{McKetta Department of Chemical Engineering, University of Texas at Austin, Austin, Texas 78712, USA}
\affiliation{Department of Physics, University of Texas at Austin, Austin, Texas 78712, USA}

\begin{abstract}
Low-density ``equilibrium'' gels that consist of a percolated, kinetically arrested network of colloidal particles and are resilient to aging can be fabricated by restricting the number of effective bonds that form between the colloids. Valence-restricted patchy particles have long served as one archetypal example of such materials, but equilibrium gels can also be realized through a synthetically simpler and scalable strategy that introduces a secondary linker, such as a small ditopic molecule, to mediate the bonds between the colloids. Here, we consider the case where the ditopic linker molecules are low-molecular-weight polymers and demonstrate using a model colloid--polymer mixture how macroscopic properties such as the phase behavior as well as the microstructure of the gel can be designed through the polymer molecular weight and concentration. The low-density window for equilibrium gel formation is favorably expanded using longer linkers, while necessarily increasing the spacing between all colloids. However, we show that blends of linkers with different sizes enable wider variation in microstructure for a given target phase behavior. Our computational study suggests a robust and tunable strategy for the experimental realization of equilibrium colloidal gels.
\end{abstract}

\maketitle

\section{Introduction}
Gels consisting of percolated, porous, and kinetically arrested networks of colloidal particles connected by chemical or physical bonds~\cite{Zaccarelli2007,Sciortino2017} possess a remarkable combination of material properties: even at extremely low particle densities, they can be self-supporting and exhibit solid-like mechanical response.~\cite{Mewis2011,Zaccarelli2007,Sciortino2017} Low density colloidal dispersions are less prone to aggregation \cite{Russel:1989} and more likely to be able to kinetically access deliberately assembled gel structures. In addition, many unique characteristics of nanoparticle dispersions are strongly perturbed in dense arrangements \cite{ElSayed:2001vv}, motivating the development of strategies to control and limit their density even while assembling them. For example, coupling interactions between plasmonically active metal nanoparticles cause spectral shifts and broadening. Controlling the extent of such interactions in gels could yield a wide range of structure-dependent optical responses; e.g., a low-density metal oxide nanocrystal gel was recently shown to retain localized surface plasmon resonance characteristic of isolated nanocrystals but with a spectral shift triggered by the assembly process. \cite{SaezCabezas:2018ux}.

Most routes to gelation are nonequilibrium in nature, e.g., the spinodal decomposition of attractive colloids is kinetically arrested.~\cite{Lu2008,Zaccarelli2008,Eberle2011} Such gels tend to have uncontrolled local density variations and age as their microstructures slowly relax towards equilibrium phase separation. In contrast, ``equilibrium'' gels~\cite{Teixeira2017,Sciortino2017,Almarza2011,delasHeras2011,Zaccarelli2005,Russo2011a,Rovigatti2013a,Russo2011b} that are resilient to aging can be formed when the arrested structure is characteristic of a true equilibrium state (i.e., a snapshot of a single phase). Gelation occurs when the lifetime of the effective bonds between colloids in this state significantly exceeds the observation time.~\cite{Mewis2011} By restricting the number of these bonds, the spinodal boundary for spontaneous phase separation can be suppressed, opening a window of fluid states that can gel at small colloid volume fractions.~\cite{Zaccarelli2005,Bianchi2006,Liu2009,Russo2009,Sciortino2017} This concept was originally demonstrated in Monte Carlo simulations for a model where the number of bonds between colloids was artificially constrained~\cite{Zaccarelli2005} and for an anisotropic (``patchy'') colloid model.~\cite{Bianchi2006,Liu2009,Russo2009} These computational studies motivated the discovery of several materials that can form equilibrium gels, including an anisotropic colloidal clay,~\cite{Ruzicka2011} DNA nanostars with complementary sticky ends~\cite{Biffi2013,Bomboi2015,FernandezCastanon2018}, and a dipeptide with several aromatic rings that allowed for specific $\pi$-$\pi$ stacking interactions.~\cite{Dudukovic2014}

Nevertheless, the preparation of equilibrium gels from patchy nanoparticles is challenged by difficulties in fabricating the latter with sufficient precision and in large enough quantities \cite{Pawar:2010}. This challenge motivated some of us to study an alternative route to equilibrium gelation that controls the extent of bonding between the (primary) colloids using a (secondary) linking agent.~\cite{Lindquist2016} Many examples of such linking systems have been described, including colloid--polymer mixtures,~\cite{Dickinson1991,Zhao2012,Chen2015,Luo2015,Feng2015} mixtures of oppositely charged colloids,~\cite{Chan2005,Gilchrist2005} ion-linked functionalized nanocrystals~\cite{Singh2015,Sayevich2016a,Sayevich2016b,Lesnyak2010}, DNA-linked nanoparticles \cite{Xiong:2008kx,Zhang:2013ky}, and DNA nanostar mixtures \cite{Romano2015,Locatelli:2017kg,Bomboi:2019gq}. The degree of bonding between the colloids can be restricted by appropriately tuning the amount of linker added. Such macroscopic control is more readily achieved in experiments than careful microscopic control of the number of patches on a nanoparticle and so is a promising strategy for fabricating low-density, equilibrium colloidal gels.

We previously demonstrated using computer simulations that the phase behavior of a mixture of equally sized spherical colloids and linkers could be straightforwardly controlled through the linker concentration.~\cite{Lindquist2016} However, at the time, we did not investigate how the design of the linker itself impacted gelation. Many linking agents, such as ditopic small molecules \cite{Borsley:2016un} or DNA linkers \cite{Xiong:2008kx,Bomboi:2019gq}, have synthetically tunable properties like their molecular size or flexibility that can be varied independently of the primary particle properties. The strength and range of the effective bonds formed by the linker between colloids should depend on these characteristics. We anticipate that the phase behavior, microstructure, and emergent macroscopic properties of a colloid--linker gel should be highly sensitive to the choice of linker, yielding a rich design space for tailoring the gel properties. 

In this article, we interrogate the phase behavior and microstructure of model equilibrium gels assembled from a mixture of spherical colloids and flexible-chain molecular (polymeric) linkers. Using a combination of thermodynamic perturbation theory and computer simulations (Sec.~\ref{sec:model}), we demonstrate how the linker length and concentration can be used to systematically modify the spinodal boundary for the mixture, which determines the colloid volume fractions at which equilibrium gelation can occur (Sec.~\ref{sec:results:phase}). The simulations reveal that many linkers bind both ends to the same particle or form double bonds between particles, especially at low colloid density, and this looping can inhibit percolation of the gel network. We also propose and demonstrate a strategy for simultaneously tuning the phase behavior and microstructure of the gel using blends of linkers with different lengths (Sec.~\ref{sec:results:micro}). Macroscopic properties of the gel, like its equilibrium modulus, are shown to have a concomitant dependence on the distribution of linker lengths (Sec.~\ref{sec:results:macro}).

\section{Model and Methods}
\label{sec:model}
To form an equilibrium gel, it is necessary to create a homogeneous, percolated network of linked colloids that subsequently undergoes an effective kinetic arrest.~\cite{Mewis2011,Sciortino2017} The region of state points where these conditions can be satisfied depends sensitively on the interactions between components in the mixture. The mixture may separate into two or more coexisting thermodynamic phases at low colloid concentrations or enter a single-phase, arrested state at high concentrations \cite{Zaccarelli2007,Sciortino2017}. Effective kinetic arrest is usually only achieved when thermal energy is sufficiently low (relative to bond strength) that the average lifetime of a bond in the percolated network becomes comparable to the experimental observation time; this point is typically gauged relative to the highest temperature for which phase separation occurs. Thus, it is important to determine the thermodynamic phase boundaries governing these transitions and to assess how they depend on the tunable linker properties.

To this end, we studied a simple model (Sec.~\ref{sec:model:model}) for a colloid--linker mixture that captures these salient physical behaviors using a combination of molecular simulations (Sec.~\ref{sec:model:sim}) and thermodynamic perturbation theory (Sec.~\ref{sec:model:tpt}). The theoretical calculations predict equilibrium phase behavior under a reasonable set of assumptions, and these predictions can be validated by the more computationally demanding simulations. The simulations additionally provide detailed microstructural information and incorporate practical complications neglected by the assumptions of the theory.

\subsection{Model}
\label{sec:model:model}
The linkers were modeled as short linear chains of $M$ segments with diameter $\sigma$ and mass $m$, which we refer to as polymers. The segment interactions were purely repulsive to mimic good-solvent conditions for the polymer \cite{Rubinstein:2003}. The colloidal particles were represented by larger spheres of diameter $\sigma_{\rm c} = 5\,\sigma$ and mass $m_{\rm c} = 125\,m$. The interactions between colloids and with the polymer segements were also purely repulsive, corresponding to good stabilization of the colloids in the (implicit) solvent and to neglecting polymer adsorption onto the colloid surface. 

\begin{figure}
    \centering
    \includegraphics{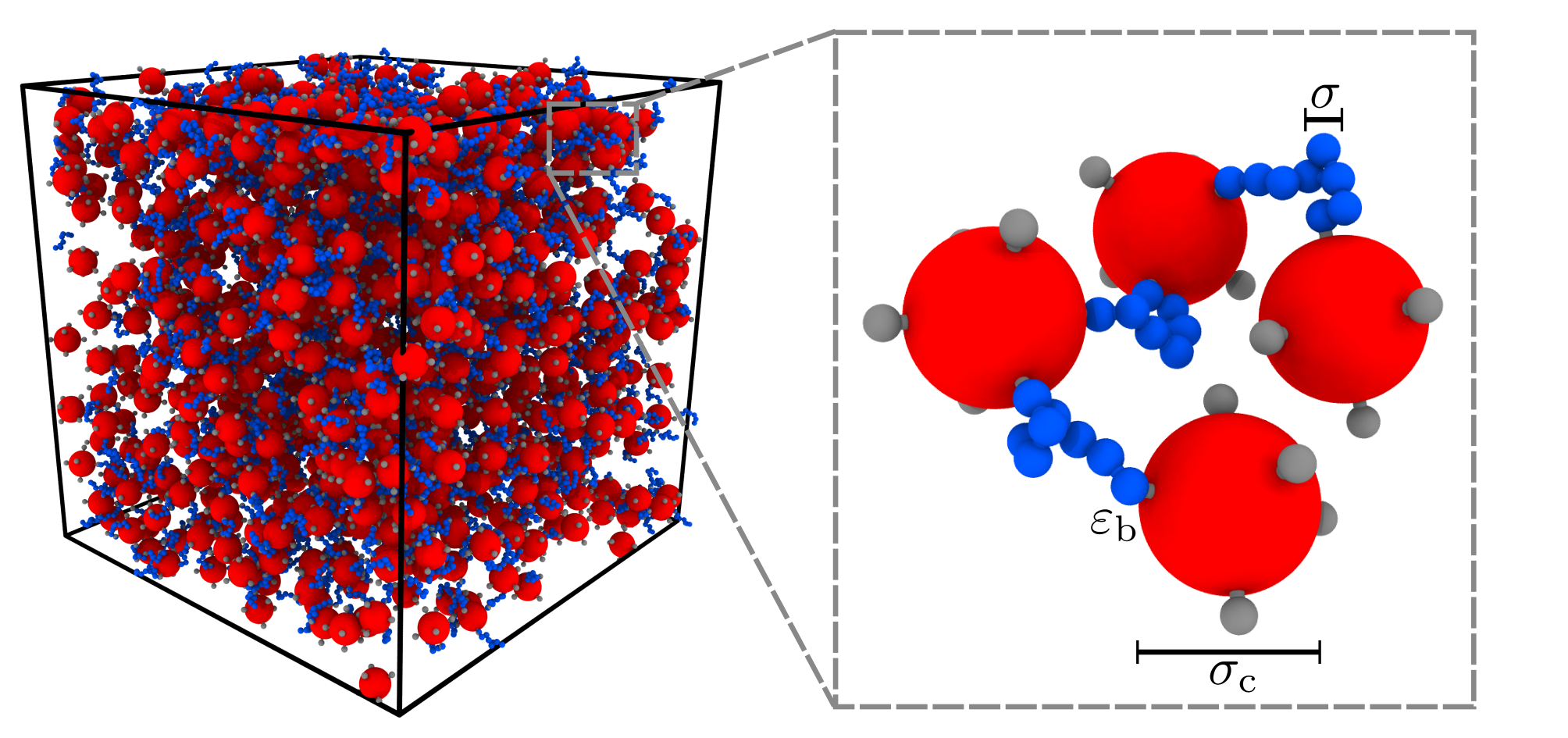}
    \caption{Snapshot of a percolated network of colloids (red) and polymers of length $M=8$ (blue) at $\Gamma = 1.5$, $\eta_{\rm c} = 0.10$, and $\beta \varepsilon_{\rm b} = 20$. The inset shows a small region of the snapshot indicating the polymer-segment diameter $\sigma$, the colloid diameter $\sigma_{\rm c}$, and the bond energy $\varepsilon_{\rm b}$ between a polymer and a colloid patch (gray). The images were rendered using OVITO 2.9.0 \cite{Stukowski:2010ky}.}
    \label{fig:snapshot}
\end{figure}

Polymer-mediated linking between colloids was incorporated by decorating the colloid surface with $n_{\rm c} = 6$ attractive patches with nominal diameter $\sigma$ and mass $m$ in an octahedral arrangement (Fig.~\ref{fig:snapshot}). The end segments of the polymer chains had a short-ranged attraction to these patches to mimic bond formation. In prior work that investigated the effect of limiting the number of bonds per particle using a single-component square-well fluid, it was shown that the phase behavior of particles that can form at least six bonds is similar to when the number of bonds per particle is not constrained.~\cite{Zaccarelli2005} Additional details about the patch model and interaction are given in Sec.~\ref{sec:model:sim}.

The colloid--polymer mixture is characterized by the temperature $T$, the volume $V$, and the number of colloids $N_{\rm c}$ and polymers $N_{\rm p}$. The total number density of the mixture is $\rho = (N_{\rm c} + N_{\rm p})/V$, and the number density for each species is $\rho_i = x_i \rho$, where $x_i = N_i/(N_{\rm c}+N_{\rm p})$ is the mole fraction for component $i$. It is convenient to describe the composition using the colloid volume fraction $\eta_{\rm c} = \rho_{\rm c} \pi \sigma_{\rm c}^3/6$ and the number ratio of polymers to colloids, $\Gamma = N_{\rm p}/N_{\rm c}$. Figure \ref{fig:snapshot} shows an example snapshot of the full model, including the colloids with their patches and the polymer linkers, for $\eta_{\rm c} = 0.10$ and $\Gamma = 1.5$.

\subsection{Simulations}
\label{sec:model:sim}
We performed molecular dynamics simulations of the colloid--polymer mixtures at constant temperature $T$ using \textsc{lammps} (16 Mar 2018).~\cite{Plimpton1995} The interparticle interactions were modeled using a core-shifted Weeks--Chandler--Andersen potential \cite{Weeks:1971uq},
\begin{equation}
    u_{\rm r}(r) = \begin{cases}
    4 \varepsilon \left[\left(\dfrac{\sigma}{r-\delta} \right)^{12} -\left(\dfrac{\sigma}{r-\delta} \right)^6 \right] + \varepsilon, & r \le r^\ast \\
    0, & r > r^\ast
    \end{cases},
    \label{eqn:wca}
\end{equation}
where $r$ is the distance between the centers of two particles $i$ and $j$, $\varepsilon = k_{\rm B} T$ sets the energy scale of the repulsion with $k_{\rm B}$ being Boltzmann's constant, and $\delta = (\sigma_i + \sigma_j)/2 - \sigma$ shifts the divergence of the potential to account for the particle diameters, $\sigma_i$ and $\sigma_j$. This potential was truncated at $r^\ast = 2^{1/6} \sigma + \delta$ to make the interactions purely repulsive and approximately hard-sphere like.

Bonds between polymer segments were modeled using finitely extensible nonlinear elastic springs \cite{Bishop:1979um},
\begin{equation}
    u_{\rm p}(r) = \begin{cases}
    -\dfrac{k_{\rm p} r_0^2}{2} \ln\left[1 - \left(\dfrac{r}{r_0} \right)^2 \right], & r \le r_0 \\
    \infty, & r > r_0
    \end{cases},
\end{equation}
with the standard Kremer--Grest parameters for the spring constant ($k_{\rm p} = 30\,k_{\rm B}T/\sigma^2$) and maximum bond extension ($r_0 = 1.5\,\sigma$) \cite{Grest:1986wy}. These parameters correspond to a polymer in a good solvent and ensure that there is no artificial chain crossing; the average bond length was $0.97\,\sigma$.

Each colloid and its patches were modeled as a rigid body. The patches were placed in an octahedral arrangement on the colloid surface at a radial distance $r^\ast$ from the center of the colloid so that a segment bonding with the patch was not repelled by the colloid ($\delta = 2\,\sigma$). The end segments of the polymer chains were attracted to these patches through a short-ranged potential,
\begin{equation}
    u_{\rm b}(r) = \begin{cases}
    -\varepsilon_{\rm b} e^{-(r/0.2\,\sigma)^2}, & r \le 0.5\,\sigma \\
    0, & r > 0.5\,\sigma
    \end{cases},
    \label{eqn:ub}
\end{equation}
where $\varepsilon_{\rm b}$ sets the bond strength. The length scale of attraction for $u_{\rm b}$ was chosen so that two polymer segments could not bond to the same patch. We additionally allowed the patches to interact with each other through eq.~\eqref{eqn:wca} to ensure that multiple patches could not coordinate with the same end segment. This computationally convenient scheme mimics the formation of an exclusive covalent bond between the patch and a segment, especially as $\varepsilon_{\rm b}$ becomes large compared to the thermal energy $k_{\rm B}T$, which is the limit relevant to typical experimentally realizable linker chemistries \cite{Borsley:2016un,Bomboi:2019gq,Reuther:2018dr}.

We prepared configurations containing $N_{\rm c} = 1000$ colloids and $N_{\rm p} = 1500$ polymers ($\Gamma = 1.5$) in a cubic, periodic simulation box of edge length $L$ at colloid volume fractions ranging from $\eta_{\rm c} = 0.01$ to 0.15. We initially equilibrated each mixture for $0.5\times 10^4\,\tau$ without any patch bonding ($\beta \varepsilon_{\rm b} = 0$), where $\tau = \sqrt{\beta m \sigma^2}$ is the unit of time and $\beta = 1/k_{\rm B}T$. The integration timestep was $0.001\,\tau$, and the temperature was controlled by a Langevin thermostat with friction coefficient $0.1\,m/\tau$ applied to each particle. We then followed a slow annealing protocol to switch on the patch attractions in an attempt to produce equilibrium gel structures. For each volume fraction, the bond energy was first linearly ramped to $\beta \varepsilon_{\rm b} = 10$ over $10^4\,\tau$, followed by a $10^4\,\tau$ equilibration period. The bond energy was then increased by $1.0\,k_{\rm B} T$ over a $0.5\times 10^4\,\tau$ period, and the mixture was allowed to reequilibrate for $1.5\times 10^4\,\tau$. The final configuration from this period was saved, and the incrementing process was repeated until $\beta \varepsilon_{\rm b} = 20$. We took the configurations generated for each $(\eta_{\rm c},\varepsilon_{\rm b})$ pair and used them to sample the structure of the mixture during production simulations of $10^4\,\tau$, saving configurations every $10\,\tau$ for subsequent analysis.

\subsection{Theory}
\label{sec:model:tpt}
Wertheim's thermodynamic perturbation theory (TPT)~\cite{Jackson1988,Chapman1988,Fantoni2015} is unrivaled in its ability to predict the equilibrium phase behavior of strong, short-ranged, highly directional, bond-forming interactions in particle-based systems. Its predictions are often in close qualitative and quantitative agreement with exact simulations of patchy particle models~\cite{Lindquist2016,delasHeras2011,Russo2011a,Rovigatti2013a,Russo2011b,Teixeira2017,Bianchi2006}. Formally, TPT is derived from a graph-theoretic expansion and a partial re-summation centered around a few key physical assumptions~\cite{Jackson1988,Chapman1988,Fantoni2015,Tavares2010}: (1) each patch on a particle can only bond with one other patch, (2) bond formation is statistically uncorrelated between patches on a particle, and (3) bonding occurs in a tree-like structure (i.e., closed loops are not permitted). As a result of these assumptions, TPT has a relatively simple closed-form expression for the free energy, rendering thermodynamic stability and equilibrium phase coexistence calculations computationally inexpensive.

Within TPT, the Helmholtz free energy $a$ is decomposed into a reference contribution $a_0$, devoid of patchy interactions, and a bonding contribution $a_{\rm b}$ that adds in the effect of patches,
\begin{equation}
a = a_0 + a_{\rm b}. \label{eqn:a}
\end{equation}
Here, the free energies are made intensive by the total number of colloids and polymers, $N = N_{\rm c} + N_{\rm p}$. We approximate the reference free energy $a_0$ as that of a hard-chain mixture assembled from a dissociated hard-sphere fluid~\cite{Chapman1988}. The linear chains of species $i$ are made up of $M_{i}$ hard spheres with diameter $\sigma_{i}$. The free energy, itself derived using TPT, is
\begin{equation} \label{eqn:a_r}
a_0 = a_{\rm hs} - \sum_i x_{i}(M_{i}-1) \text{ln}g_{ii}(\sigma_i^+),
\end{equation}
where $a_{\rm hs}$ is the free energy of an equivalent hard-sphere mixture obtained by fully dissociating all chains, and $g_{ii}(\sigma_i^+)$ is the corresponding contact value for the radial distribution function of species $i$ in the hard-sphere fluid. Practically, we employ Boubl\'{i}k's well-known and accurate equation of state \cite{Boublik:1970um} for $a_{\rm hs}$ and $g_{ii}(\sigma_i^+)$.

The bonding contribution to the free energy is
\begin{equation} \label{eqn:a_b}
a_{\rm b} = \sum_i x_{i} \left[ n_{i}\text{ln}X_{i} - \dfrac{n_{i}X_{i}}{2} + \dfrac{n_{i}}{2}  \right],
\end{equation}
where $n_i$ is the number of patches and $X_i$ is the fraction of patches that are not bonded for species $i$. The values of $X_i$ are determined by a coupled set of chemical equilibrium equations,
\begin{equation} \label{eqn:Xi}
X_{i} = \left[ 1 + \rho \sum_j n_{j}x_{j} X_{j} \Delta_{ij} \right]^{-1},
\end{equation}
with $\Delta_{ij}$ being an effective bond partition function. For our model, where only polymers bond to colloids,
\begin{equation} \label{eqn:Delta}
\Delta_{\rm cp} \approx \int {\rm d}\mathbf{r} g_{\rm cp}(r) \langle f_{\rm b} | r \rangle,
\end{equation}
which depends on both the radial distribution function $g_{\rm cp}$ between the colloid (c) and polymer segment (p) and the bonding Mayer f-function \cite{Hansen2013,Jackson1988,Chapman1988} $f_{\rm b}$ averaged over all angular orientations of the colloid and polymer segment at a center-to-center distance $r = |\mathbf{r}|$. As a further simplification, we assume that the range where $f_{\rm b}$ is nonzero is short and take $g_{\rm cp}(r) \approx g_{\rm cp}(\sigma_{\rm cp}^+)$ for a reference hard-sphere fluid at contact distance $\sigma_{\rm cp} = (\sigma_c + \sigma)/2$. Details regarding the evaluation of eq.~\eqref{eqn:Delta} are provided in the Appendix.

To divide parameter space into homogeneous (single-phase fluid) and phase-separated regimes, we computed the spinodal and binodal boundaries for the mixture. The spinodal demarcates the conditions under which the single-phase fluid becomes unstable towards spontaneous decomposition into multiple phases; true equilibrium phase separation at the binodal always preempts the spinodal, and the single-phase fluid is only metastable in the region between the binodal and spinodal~\cite{Chimowitz2005,Shell2015}. At fixed $N$, $V$, and $T$, the spinodal is determined by the locus of points for which the determinant of the stability matrix $\mathbf{H}$ is zero, with $H_{ij} = \partial^{2} a / \partial x_{i} \partial x_{j}$ being the second derivative of the Helmholtz free energy with respect to composition~\cite{Chimowitz2005,Shell2015}. The binodal is found by postulating a two-phase system and minimizing its total free energy subject to particle and volume conservation between phases; the binodal corresponds to the conditions where an infinitessimal amount of a new phase emerges. Practical details regarding this approach are available elsewhere~\cite{Lindquist2016}.

\section{Results and Discussion}
\label{sec:results}

\subsection{Phase behavior and linker length}
\label{sec:results:phase}
We first investigated the phase behavior of mixtures of colloids with linkers of a single length. Unlike our prior work with spherical linkers,~\cite{Lindquist2016} the polymer linkers are flexible and possess internal degrees of freedom, leading to softer effective interactions between linked colloids. Because the linkers in the prior work occupied the same excluded volume as the primary colloids, the polymer linkers considered here also occupy a lower volume fraction. Hence, we hypothesized that the colloid--polymer mixtures might exhibit different thresholds for network percolation and phase separation than the sphere mixtures, and that the phase behavior may depend on the polymeric linker length.

The formation of a percolated, self-supporting network is the first requirement for gelation. Percolation only occurs once a sufficient amount of linker has been added at cold enough temperature (high enough bond energy) and can be predicted using TPT as the conditions under which there is unbounded growth of the network of effective bonds between colloids. In the limit $\beta\varepsilon_{\rm b} \to \infty$, relevant to strong bonds between the linkers and colloids, the percolation threshold depends simply on the patch stoichiometry and is estimated to be $\Gamma = 0.6$ for our model. In previous simulations, though, we previously found that percolation did not occur until at least $\Gamma = 0.9$ for the spherical linkers.~\cite{Lindquist2016}

Based on this, we initially performed simulations at $\Gamma = 1.0$, but were unable to form a percolated network even at colloid volume fractions predicted to be outside the phase separated region. In the simulations, percolation was defined as the formation of a network spanning the periodic simulation box in at least one dimension. We speculated that this discrepancy was caused by linkers forming ``loops'' in the colloid network, which are neglected by TPT. Redundant linker bonds create cycles in the network graph having length equal to the number of linkers in the loop, and lead to a smaller number of effective connections between colloids than expected. When we increased the linker concentration to $\Gamma = 1.5$, we were able to obtain a percolated network.

\begin{figure}
    \centering
    \includegraphics{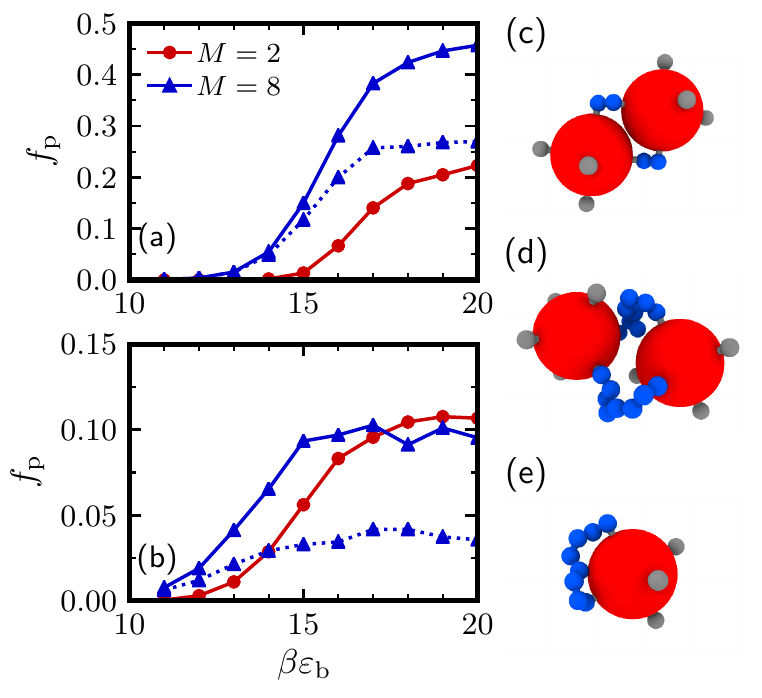}
    \caption{Fraction of polymers, $f_{\rm p}$, that formed redundant bonds in the network as a function of bond energy, $\varepsilon_{\rm b}$, at $\Gamma = 1.5$. The solid lines indicate both self bonds (cycle length 1) and one of the polymers in the double bonds (cycle length 2), while the dashed line indicates only those from self bonds. The colloid volume fractions are (a) $\eta_{\rm c} = 0.01$, which phase separated, and (b) $\eta_{\rm c} = 0.15$, which formed a homogeneous, percolated network. Note the difference in scale for $f_{\rm p}$ between (a) and (b),
    as there was more significant looping at smaller $\eta_{\rm c}$. (c) and (d) show snapshots of the double bond for $M=2$ and $M=8$, while (e) shows the self bond for $M=8$.}
    \label{fig:loops}
\end{figure}

We can characterize the number of linkers that do not form new connections in the colloid network directly from these simulations. We considered the two shortest looping motifs, which we expected to occur in largest number and so to have the greatest impact on the phase behavior of the system: a cycle of length one where a linker bonds back onto the same colloid, and a cycle of length two, where two linkers form a double bond between two colloids. We counted the fraction of linkers $f_{\rm p}$ that formed redundant bonds from these motifs for linker lengths $M=2$ and 8 at $\Gamma = 1.5$ as a function of $\varepsilon_{\rm b}$ (Fig.~\ref{fig:loops}). Only the $M=8$ linkers were long enough to bond two patches on the same colloid, but both linkers were able to form double bonds. We considered both low ($\eta_{\rm c} = 0.01$) and high ($\eta_{\rm c} = 0.15$) colloid densities, which TPT predicted to be phase separated and homogeneous, respectively.

For low $\varepsilon_{\rm b}$, there were a negligible number of redundant bonds in the network, although the total number of bonds was also lower. As $\varepsilon_{\rm b}$ increased, $f_{\rm p}$ slowly increased until a sharp rise occurred near $\beta\varepsilon_{\rm b} \approx 15$. This rise in $f_{\rm p}$ happened at lower $\varepsilon_{\rm b}$ for the $M=8$ linkers, where most of these redundant bonds initially came from linkers bonded to the same colloid. The total number of redundant bonds at $\beta \varepsilon_{\rm b} = 20$ was much larger in the low density mixture (20\% to 40\% of linkers) than in the high density mixture (10\% of linkers). However, because $f_{\rm p}$ was similar for both mixtures when $\beta \varepsilon_{\rm b}$ was smaller, this increased tendency to form redundant bonds at large $\varepsilon_{\rm b}$ for lower $\eta_{\rm c}$ may be due to the onset of phase separation. The redundant bonds likely inhibited percolation compared to TPT predictions, and are a real effect that must be considered in an experimental realization of linker gels.

We then determined approximate phase diagrams of the colloid--polymer mixtures from the simulations in order to compare with TPT's predictions. It is challenging to rigorously compute these diagrams from the free energies of competing phases, and so we instead looked for signs of phase separation emerging within our finite simulation box and finite observation time. We computed the partial static structure factor for the colloids \cite{Hansen2013},
\begin{equation}
    S_{\rm cc}(\mathbf{q}) = \frac{1}{N_{\rm c}} \left\langle \sum_{j,k}^{N_{\rm c}} e^{-i \mathbf{q} \cdot (\mathbf{r}_j - \mathbf{r}_k)} \right\rangle,
\end{equation}
where $\mathbf{q} = 2\pi \mathbf{n}/L$ is the wavevector, $\mathbf{n}$ is a vector of integers, and $\mathbf{r}_j$ is the position of the $j$-th colloid. We determined $S_{\rm cc}$ for the 22 smallest nonzero wavevector magnitudes $q = |\mathbf{q}|$, averaging over the wavevectors corresponding to each $q$. We then extrapolated the structure factor to zero-wavevector, $S_{\rm cc}(0)$, by fitting $S_{\rm cc}$ to a Lorentzian form \cite{Jadrich:2015jb},
\begin{equation}
    S_{\rm cc}(q) \approx \frac{S_{\rm cc}(0)}{1 - (q \xi)^2},
\end{equation}
with $\xi$ being a correlation length. In the thermodynamic limit, $S_{\rm cc}(0)$ should diverge at the spinodal, but in practice, it remains finite in the simulation and $S_{\rm cc}(0) > 10$ is usually considered to be phase separated.~\cite{Zaccarelli2005,Lindquist2016} We qualitatively confirmed that this criterion held for our model by visually inspecting configurations that it identified as phase separated, finding that they had clear density variations.

\begin{figure}
    \centering
    \includegraphics{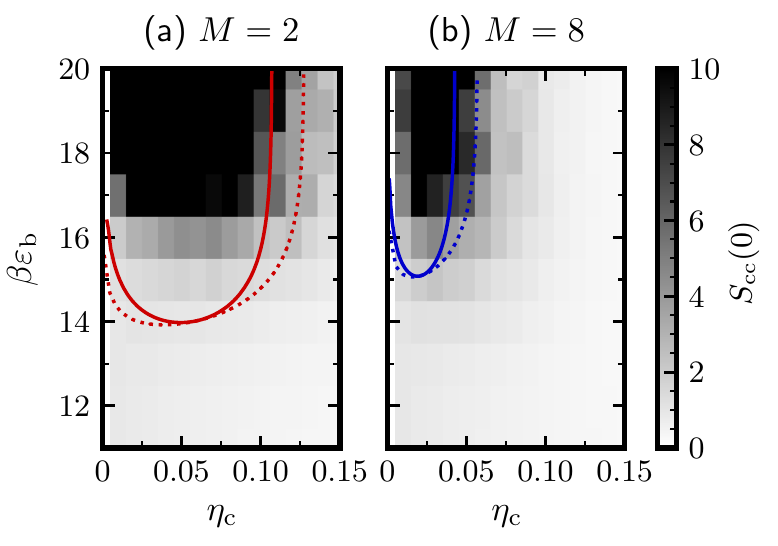}
    \caption{Phase diagrams for colloids mixed with polymers of length (a) $M = 2$ and (b) $M = 8$ in the $(\eta_{\rm c},\varepsilon_{\rm b})$ plane at $\Gamma = 1.5$. The solid lines indicate the spinodals computed from TPT, while the dotted lines show the binodals. The grayscale map gives the colloid partial structure factor extrapolated to zero-wavevector, $S_{\rm cc}(0)$, from the simulations. Large values of $S_{\rm cc}(0)$ indicate phase separation in the simulations; usually $S_{\rm cc}(0) > 10$ is considered separated.}
    \label{fig:pdone}
\end{figure}

We compared the measured structure factor to the TPT predictions of phase coexistence for two chain lengths, $M=2$ and $M=8$, with $\Gamma = 1.5$ (Fig.~\ref{fig:pdone}). The theoretical spinodal boundary (solid line) indicates the state points for which the system will spontaneously demix, while the binodal boundary (dotted line) fully encompasses the region where phase separation can occur, including metastable regions. Phase separation should nearly always occur in the simulations for state points inside the spinodal, but may or may not be observed in the metastable region between the binodal and spinodal due to the finite simulation time.

Overall, the simulations and TPT were in good agreement, particularly with respect to $\eta_{\rm c}$. Accurately predicting the phase boundary with respect to $\eta_{\rm c}$ is important for determining the colloid volume fractions that can form a low-density equilibrium gel. We additionally confirmed that the structures that were not phase separated formed percolated networks once $\varepsilon_{\rm b}$ was sufficiently large, typically around $\beta\varepsilon_{\rm b} \gtrsim 16$ for $\eta_{\rm c} = 0.15$. There was a discrepancy of about 15\% between the simulations and TPT in the minimum $\varepsilon_{\rm b}$ needed for phase separation to occur. We speculate that this is due to the presence of linker loops in the colloid network. The discrepancy is qualitatively consistent with the number of redundant bonds in the network near the onset of phase separation (Fig.~\ref{fig:loops}), but it is not obvious how to renormalize the TPT to account for this or cycles of longer lengths. Ultimately, reliably predicting the minimum $\varepsilon_{\rm b}$ needed for phase separation is not as crucial for experimental realization of the gels as predicting the minimum $\eta_{\rm c}$ at which single phase gel assemblies may be realized. TPT can be reasonably used to qualitatively assess trends in the phase behavior as functions of linker length and mixture composition and to suggest experimental strategies to realize such colloidal gels at low density.

\begin{figure}
    \centering
    \includegraphics{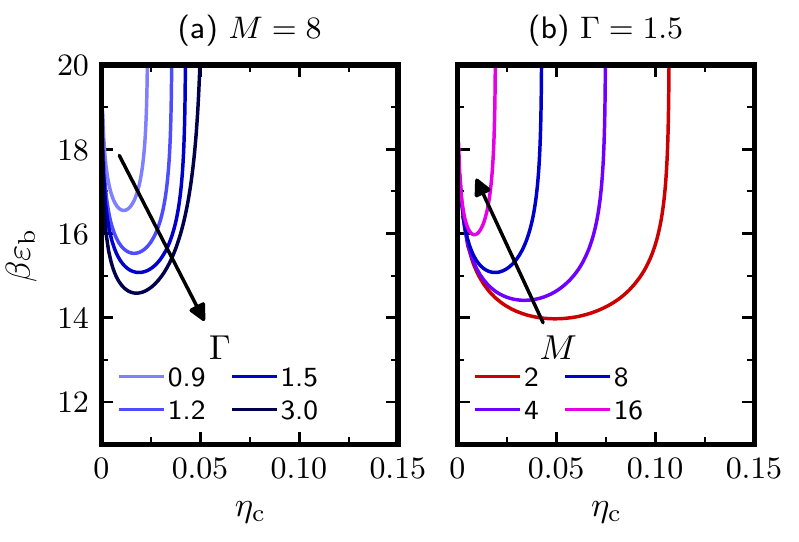}
    \caption{Spinodal boundaries predicted by TPT in the $(\eta_{\rm c}, \varepsilon_{\rm b})$ plane for (a) polymers of length $M=8$ with varied $\Gamma$ and (b) polymers of different length $M$ with $\Gamma = 1.5$. Note that the spinodal boundary shrinks dramatically as $M$ increases but only broadens weakly with increasing $\Gamma$.}
    \label{fig:pdtheory}
\end{figure}

In our previous work \cite{Lindquist2016}, we showed how $\Gamma$ controls the onset of percolation and also the range of colloid volume fraction over which phase separation is expected for colloids joined by spherical linkers. These mixtures exhibited reentrant phase behavior as a function of $\Gamma$, as the spinodal initially broadened after percolation with a maximum width at the stoichiometric ratio of linkers with the 6-patch colloids ($\Gamma = 3$), before shrinking and eventually vanishing at higher $\Gamma$. In this work, we are primarily interested in the low-linker ($\Gamma \le 3$) regime, where the phase behavior is not as strongly influenced by the number of bonding sites on the colloids. Moreover, restricting the linker concentration may help circumvent linker solubility issues \cite{Rubinstein:2003} and avoids the introduction of additional interactions to the system (e.g., depletion attraction \cite{Asakura:1954ts,Russel:1989} between colloids due to free linkers \cite{SaezCabezas:2018ux}) that may affect the phase behavior. For polymeric linkers with $M=8$, analogous to the spherical linkers, the spinodal broadens as $\Gamma$ is increased from 0.9 to 3.0 (Figure \ref{fig:pdtheory}). However, the shift in the high density branch of the spinodal with respect to $\eta_{\rm c}$ is only a factor of 2.1. Moreover, $\Gamma$ is additionally constrained by the need to form a percolated network and so can be decreased over only a limited range.

In contrast, the spinodal boundary is far more sensitive to the length of the linkers. Figure \ref{fig:pdtheory}b shows that the high-density branch of the spinodal can be shifted by roughly a factor of 5.6 with respect to $\eta_{\rm c}$ as $M$ is increased from $M=2$ to $M=16$. This trend is also born out in the simulations (Fig.~\ref{fig:pdone}). In patchy particle systems, an analogous compression of the spinodal is achieved by limiting the number of patches on the particles, but here we are able to keep the primary particles the same and simply adjust the linker length. These results demonstrate that the linker length can be used to control and expand the low-density window for producing equilibrium gels. 

\subsection{Microstructure and linker mixtures}
\label{sec:results:micro}
Although linker length provides effective control over the phase behavior of the system, it is expected to simultaneously affect the microstructure of the gel. Longer polymers occupy and exclude a larger volume than shorter polymers, which should increase the average distance between the colloids they connect; a similar effect has been observed in DNA-linked colloidal crystals \cite{Nykypanchuk:2008ea,Hill:2008kz}. This increased interparticle spacing may have consequences for the macroscopic properties of the gel. For example, near-field coupling between plasmonic nanoparticles rapidly decays with increased surface separation \cite{Jain:2007}. Hence, it may be necessary to simultaneously consider the phase behavior and microstructure of the gel when designing the linker.

\begin{figure}
    \centering
    \includegraphics{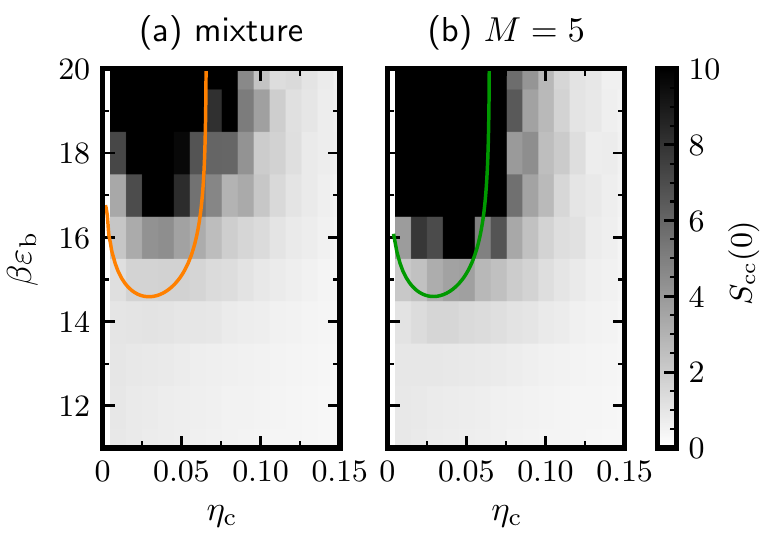}
    \caption{Phase diagrams in the $(\eta_{\rm c},\varepsilon_{\rm b})$ plane at $\Gamma = 1.5$ and for colloids mixed with (a) an equimolar mixture of $M=2$ and $M=8$ polymers and (b) polymers of length $M = 5$. The grayscale map shows the colloid partial structure factor, determined from simulations as in Fig.~\ref{fig:pdone}. The spinodal boundaries predicted by TPT are highly similar for both (a) and (b).}
    \label{fig:pdmix}
\end{figure}

Given that the average interparticle spacing for a given linker is likely intimately related to its length, we propose one strategy to partially decouple the phase behavior and microstructure by using mixtures of linkers having different lengths. The linker mixture is expected to lead to different phase behavior and microstructure than its pure components, and these should both be sensitive to the linker lengths and concentrations. To demonstrate this idea, we considered an equimolar mixture of $M=2$ and $M=8$ linkers. We computed the spinodal boundary for this system using TPT and also ran equivalent simulations (Fig.~\ref{fig:pdmix}a). The spinodal of the mixture lies in between that of the systems with only $M=2$ or $M=8$ linkers (Fig.~\ref{fig:pdone}), and the TPT prediction was again in reasonable agreement with the simulations. We further compared the phase behavior of this mixture to an effective single-linker system having the same number average molecular weight, $M=5$ (Fig.~\ref{fig:pdmix}b). Interestingly, the spinodal predicted by TPT is nearly the same for the mixture and the effective linker, and the phase behavior measured in the simulations shows similar boundaries in $\eta_{\rm c}$ with some discrepancies again arising in $\varepsilon_{\rm b}$.

\begin{figure}
    \centering
    \includegraphics{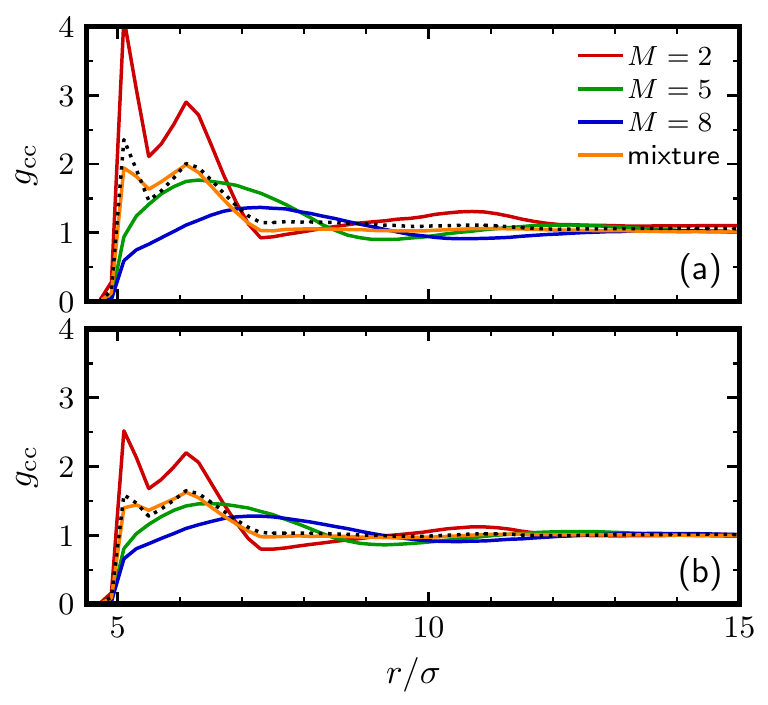}
    \caption{Colloid radial distribution function, $g_{\rm cc}$, at (a) $\eta_{\rm c} = 0.10$ and (b) $\eta_{\rm c} = 0.15$ with $\beta \varepsilon_{\rm b} = 20$ and $\Gamma = 1.5$ for polymers of lengths $M = 2$, 5, and 8, and an equimolar mixture of $M=2$ and $M=8$ polymers. The dotted line indicates the average of the measured $g_{\rm cc}$ for $M=2$ and $M=8$.}
    \label{fig:rdf}
\end{figure}

Despite this similarity in phase behavior, the simulations revealed that the linker mixture had a significantly different microstructure than the single linker with the same average length. We computed the radial distribution function $g_{\rm cc}$ between the colloids for the three single-linker systems studied ($M=2$, 5, and 8) and also the mixture at different $\eta_{\rm c}$ (Fig.~\ref{fig:rdf}). As the linker length increased from $M=2$ to $M=8$, the first peak in $g_{\rm cc}$ was pushed outward to larger $r$, from near contact at $r \approx 5\,\sigma$ for $M=2$ to $r \approx 7\,\sigma$ for $M=8$. This increase in the colloid spacing with linker length was expected because the preferred size of a linear chain increases with $M$ (polymer radius $R \sim M^{0.588}$ in a good solvent), and there is a corresponding free energy penalty to confine the linkers between two closely-spaced colloid surfaces \cite{Rubinstein:2003}.

The microstructure of the mixture was dissimilar to the equivalent single-linker ($M=5$), instead showing signatures of particle spacings typical of both the $M=2$ and $M=8$ component linkers and a more uniform long-ranged structure on average. The split peaks close to contact, which also arise for the $M=2$ linkers, are a key indicator of the underlying local heterogeneity of the microstructure in the mixture. These peaks are due to colloids in configurations rotated around the short $M=2$ (dimer) bond, from contact to separation by the bond length, and are absent for the longer linkers.

We measured $g_{\rm cc}$ at both $\eta_{\rm c} = 0.10$ (Fig.~\ref{fig:rdf}a) and $\eta_{\rm c} = 0.15$ (Fig.~\ref{fig:rdf}b), obtaining the same qualitative trends for the first peaks in $g_{\rm cc}$ at both volume fractions. $g_{\rm cc}$ was largely unchanged by $\eta_{\rm c}$ for the $M=8$ linkers, which is consistent with a sufficiently low-density fluid interacting through a potential of mean force determined by the linker. As the linker length decreased, the first peaks in $g_{\rm cc}$ became more pronounced at the lower $\eta_{\rm c}$. This additional structuring is likely due to the approach to the spinodal boundary for the $M=5$ linkers and the mixture (Fig.~\ref{fig:pdmix}), where increased long-ranged structuring was also detected by $S_{\rm cc}(0)$. In fact, at $\eta_{\rm c} = 0.10$, the $M=2$ linker has crossed into the spinodal region (Fig.~\ref{fig:pdone}a).

These results demonstrate how the linker composition can be used to design the phase behavior and microstructure of colloidal gels at a target volume fraction. The $M=2$ linkers promote a microstructure with close contacts between particles, but the colloid-linker mixture phase separates at $\eta_{\rm c} = 0.10$. The $M=5$ and $M=8$ linkers can form homogeneous equilibrium gels at $\eta_{\rm c} = 0.10$, but the first neighbors of the colloids are farther apart. By blending the $M=2$ and $M=8$ linkers, an equilibrium gel can be produced at $\eta_{\rm c} = 0.10$ with a microstructure that shows signatures of both linkers. In fact, $g_{\rm cc}$ for the mixture was effectively given by a composition-weighted average of $g_{\rm cc}$ for the pure single linker fluids (dotted line in Fig.~\ref{fig:rdf}).

Using a mixed linker composition not only provides facile control over gelation through the linker concentration, but also offers flexibility in extending the low-density window for gelation and tuning the microstructure of the resulting gel; this paradigm can be applied to other linker systems with a broad range of linking chemistries. Moreover, the linker gel clearly offers greater tunability compared to a single-component gels, including those based on conventional patchy particles, because linker properties like concentration, length, and flexibility can be controlled independently of the primary colloid.

\subsection{Macroscopic properties}
\label{sec:results:macro}
\begin{figure}
    \centering
    \includegraphics{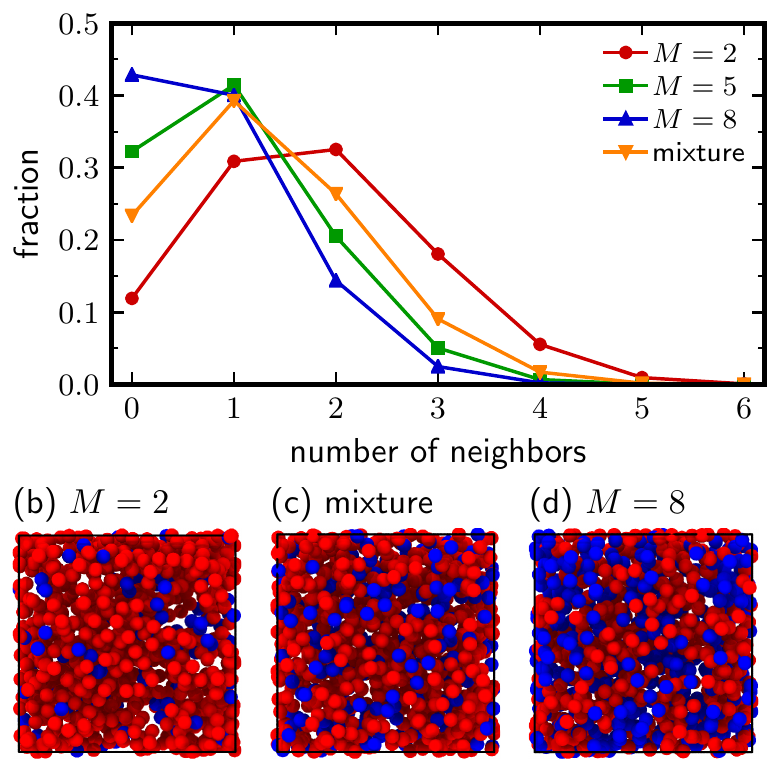}
    \caption{(a) Average fraction of colloids having a given number of neighbors with surface separation $1\,\sigma$ (20\% of $\sigma_{\rm c}$) for $\eta_{\rm c} = 0.15$ and $\beta\varepsilon_{\rm b} = 20$. Representative snapshots of (b) $M=2$, (c) equimolar mixture, and (d) $M=8$ distinguishing particles having no such neighbors (blue) from those having at least one neighbor (red).}
    \label{fig:coord}
\end{figure}

Having established how the linker can be used to design phase behavior and microstructure of the gel, we last considered potential impacts of the linker on macroscopic properties that depend on the underlying gel structure. As a rough proxy for the impact of the linker length on the optical properties of a gel composed of plasmonically active nanoparticles, we computed the average fraction of colloids in a given configuration having a certain number of neighbors with surface separation less than $1\,\sigma$ (Fig.~\ref{fig:coord}). This separation corresponds to 20\% of $\sigma_{\rm c}$, which is the distance at which significant coupling between plasmonic nanoparticles may be expected to influence the macroscopically observable optical properties of such a gel \cite{Jain:2007}. Hence, the optical spectra of plasmonic nanoparticle gels are expected to depend strongly on the average number of closely contacted neighbors.

Consistent with Fig.~\ref{fig:rdf}, the shorter linkers created microstructures with more colloids having at least one close contact (88\% for $M=2$), while the longer linkers had far fewer of these (57\% for $M=8$), as shown in Fig.~\ref{fig:coord}a. The mixture had more colloids with close contacts (77\%) than its effective single-length linker (68\%), and its coordination was intermediate between its component linkers. The differences in coordination between the mixture and its component linkers are visually apparent in representative snapshots from the simulations (Fig.~\ref{fig:coord}b--d). This example clearly highlights an advantage of the linker mixtures compared to the single-length linker: the microstructure can be reasonably controlled while simultaneously maintaining the desired phase behavior. However, further study using chemically detailed simulations and a quantitative electromagnetic model will be required to fully understand how the linker can be used to modulate, e.g., the absorption spectrum of a gel composed of plasmonically active nanoparticles.

We also expected that the linker length might influence the gel's mechanical properties, which have important implications for how the assembled gel responds to applied stresses during use. The linkers create effective bonds between the colloids that can store energy. The gel can then be thought of as an elastic network \cite{Surve:2006} analogous to classic theories for polymers, with the primary colloids serving as effective cross-link junctions. In the polymer models, the shear modulus is dominated by entropy and so is roughly $k_{\rm B} T$ per network strand per volume \cite{Rubinstein:2003}. It was recently shown, however, that the degree of bonding controls the rheology of DNA nanostar hydrogels \cite{Conrad:2019wn} through both the elasticity of the effective bonds in the network and the constraints it imposes on the junctions themselves. Moreover, real chain molecules have additional interactions like excluded volume that prevent chain crossing, which can cause the modulus to increase \cite{Everaers:1999}. Hence, we characterized the linear rheology for our linker model directly from the simulations.

To simulate long-lived network connections relevant to experiments ($\beta \varepsilon_{\rm b}\to\infty$) without sacrificing numerical accuracy, we permanently bonded the end segments of the polymers to the colloids. (In practice, this was done in \textsc{lammps} by removing and replacing the overlapped colloid patches in the rigid body by the end segment of the polymer.) We additionally switched off the attractions to the remaining patches from any free polymer ends to completely freeze the network configuration. We then measured the equilibrium stress tensor $\bm{\sigma}$ over trajectories generated from 21 statistically independent network configurations for each linker length at $\eta_{\rm c} = 0.15$. We computed the shear-stress relaxation function from the off-diagonal ($i \ne j$) components of the stress,
\begin{equation}
    G(t) = \frac{V}{k_{\rm B}{T}} \langle \sigma_{ij}(0) \sigma_{ij}(t) \rangle,
\end{equation}
for $t < 100\,\tau$. Because of the permanent network bonds \cite{Rubinstein:2003}, $G$ did not decay to zero at long times but rather reached an equilibrium modulus $G_{\rm e}$. We determined $G_{\rm e}$ by averaging $G$ over times $t \ge 25\,\tau$.

\begin{figure}
    \centering
    \includegraphics{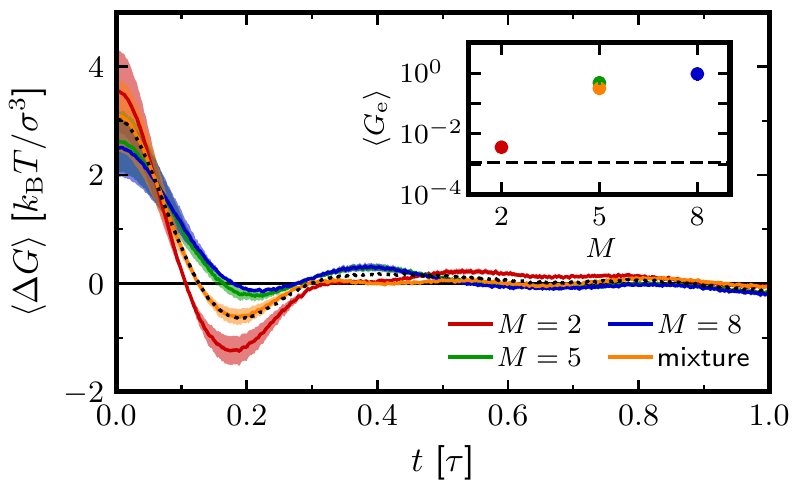}
    \caption{Average stress relaxation function $\Delta G$ relative to the equilibrium modulus $G_{\rm e}$ (inset) at $\eta_{\rm c} = 0.15$ for polymers of lengths $M = 2$, 5, and 8, and an equimolar mixture of $M=2$ and $M=8$ polymers. Error bars indicate one standard error computed from 21 statistically independent network configurations with permanent bonds and are comparable to the symbol size for the inset. The dotted line indicates the average of the measured $\Delta G$ for $M=2$ and $M=8$. The dashed line in the inset is the modulus predicted by Flory's theory of elastic networks with $\xi = (\Gamma-1)N_{\rm c}-1$.}
    \label{fig:stress}
\end{figure}

To compare the stress relaxation of the different network configurations and linker lengths, we defined $\Delta G(t) = G(t) - G_{\rm e}$ (Fig.~\ref{fig:stress}). We averaged $\Delta G$ and $G_{\rm e}$ across the independent configurations of a given linker length. There were clear differences in $\Delta G$ for the different linker lengths, with $M=2$ showing a faster initial decay in the stress followed by a stronger anticorrelation than the longer linkers. The differences between the $M=5$ and $M=8$ linker were less pronounced. Interestingly, the mixture of $M=2$ and $M=8$ linkers showed a stress relaxation intermediate between the systems with one of those two linkers. This relaxation was initially consistent with the arithmetic mean of the two, although small differences are apparent for longer times; this deviation may be partially due to statistical uncertainty in the simulations. These results suggest that linker mixture composition can then be used to tune the macroscopic effective stress relaxation in the gel, which is connected to the frequency-dependent moduli.

The equilibrium modulus $G_{\rm e}$ also systematically increased with  linker length (inset of Fig.~\ref{fig:stress}), although there was significant variability due to the different network configurations. Flory's elastic theory for cross-linked polymer networks \cite{Flory:1976ke,Flory:1985cc} predicts a modulus of $\xi (k_{\rm B} T/V)$, where $\xi = B - J - C$ is the cycle rank of a network having $B$ bonds, $J$ junctions, and $C$ connected components. If all linkers were incorporated into our network uniformly and without any defects, $\xi = (\Gamma-1)N_{\rm c} - 1$, giving a modulus of $0.001 k_{\rm B}T/\sigma^3$ at $\eta_{\rm c} = 0.15$. Clearly, $G_{\rm e}$ was much larger than this for most of the points, especially as the linker length increased. We confirmed that $\xi$ did not deviate significantly ($<5\%$) from its idealized value for any of the networks. We accordingly speculate that the large differences in $G_{\rm e}$ are due to approximations made mapping the simulated linker gel onto the elastic network model, namely, the neglect of excluded volume interactions between monomers, which may cause the polymer chains to deviate from the theoretically assumed Gaussian statistics and prevent chain crossing, and from the large colloids that form the network junctions, which have both excluded volume and rotational degrees of freedom. Further work is required to quantitatively understand how and why $G_{\rm e}$ and other rheological properties depend on $M$ for colloid--polymer linker gels.

\section{Conclusions}
We have demonstrated how the phase behavior and microstructure of an equilibrium linker gel assembled from a model colloid--polymer mixture can be systematically tuned through the linker concentration and length using a combination of thermodynamic perturbation theory and computer simulations. The spinodal boundary for phase separation was dramatically compressed toward low colloid volume fractions upon increasing the linker length, predicting the potential for realizing equilibrium gels at low density. However, these low density gels consisted of colloidal particles with an increased spacing in the microstructure, which may have macroscopic consequences on the gel's optical properties or rheology. To offer flexibility in designing gels that reach targeted properties of interest, we proposed using blends of linkers of different lengths to allow independent control over both the phase behavior, which was consistent with an effective single-length linker, and the microstructure, which was consistent with an average of the constituent linkers. Together, these results constitute a robust and facile strategy for fabricating equilibrium gels with tunable densities and microstructures by straightforward macroscopic  choices that are experimentally accessible, namely, the length of polymeric linkers and their concentrations relative to the colloids undergoing assembly.

Our simulations revealed that a substantial fraction of linkers did not form effective bonds in the colloidal gel network, either due to bonding both ends to the same colloid (looping) or by forming multiple bridging linkages between the same two neighboring particles, i.e., a double bond. Both conditions represent a ``waste'' of linkers that are no longer available to contribute to percolation of the colloidal gel network. We expect that the fraction of linkers forming such bonds would increase with the number of binding patches on the  colloidal particles. Since these wasted linkers may inhibit percolation of the gel network, this effect should be taken into account in the linker design and formulation of experimental systems. For example, it may be possible to optimize the linker itself or its interactions with the colloids to limit the number of such bonds that form.

In this article, we restricted ourselves to consider only the effect of linker length on the phase behavior and microstructure, but other linker properties like internal flexibility may similarly impact the gel. Moreover, we also neglected solvent-mediated interactions \cite{Dickinson:2013dx}, including effective interactions between the linkers and the colloids and also hydrodynamic interactions between the various components. We plan to investigate these effects with a particular focus on how the linker influences macroscopic properties like the gel rheology in future work.

\begin{acknowledgments}
We thank Zachary Sherman for helpful comments on this manuscript.
This research was primarily supported by the National Science Foundation through the Center for Dynamics and Control of Materials: an NSF MRSEC under Cooperative Agreement No. DMR-1720595 and also by the Welch Foundation (Grant Nos. F-1696 and F-1848). We acknowledge the Texas Advanced Computing Center (TACC) at The University of Texas at Austin for providing HPC resources.
\end{acknowledgments}

\appendix*
\section{Evaluating the bond overlap function}
\label{appendix_a}
Without loss of generality, we place the colloid at the origin and assume that the polymer segment lies a distance $r$ along the $z$-axis. The polymer patch is effectively at the center of the segment from eq.~\eqref{eqn:ub}, and so rotation of the segment can be neglected from the average in eq.~\eqref{eqn:Delta}. The colloid patch is offset a distance $r^\ast$ from its center, which is a point on the surface of a sphere with coordinates $(r^*,\theta,\phi)$, where $0 \le \theta < 2\pi$ is the azimuthal angle and $0 \le \phi < \pi$ is the polar angle. The vector pointing to the colloid patch from the polymer segment is $\Delta \mathbf{r} = [r^\ast \cos{\theta} \sin{\phi}, r^\ast \sin{\theta}\sin{\phi}, r^\ast \cos{\phi}-r]$. The average of the Mayer f-function is then
\begin{equation} \label{eqn:f_avg_explicit}
\langle f_{\rm b} | r \rangle = \frac{1}{4\pi} \int_{0}^{2\pi} {\rm d}\theta \int_{0}^{\pi}{\rm d}\phi \sin{\phi} \left( e^{-\beta u_{\rm b}(|\Delta \mathbf{r}|)}-1 \right),
\end{equation}
where the integrand depends on both $\theta$ and $\phi$ through $\Delta \mathbf{r}$; this integral is easily computed by quadrature. However, for efficiency in the spinodal and binodal calculations that require many evaluations of this expression, we first evaluated eq.~\eqref{eqn:f_avg_explicit} at numerous values of $\beta \varepsilon_{\rm b}$ between 0 and 20 and then fit the entire functional form to a basic neural network model that produced virtually exact results over the fit range.

\bibliography{ms}

\end{document}